\documentclass{ifacconf}
\usepackage{graphicx}      
\usepackage{color}
\usepackage{epstopdf}
\usepackage{comment}
\usepackage[square, comma, sort&compress, numbers]{natbib}
\usepackage{amsmath} 
\usepackage{amssymb}  

\newtheorem{proposition}{Proposition}

\newtheorem{corollary}{Corollary}
\newtheorem{definition}{Definition}

\newtheorem{example}{Example}

\setcitestyle{square}

\begin{document}

\begin{frontmatter}

\title{Timed Discrete-Event Systems are Synchronous Product Structures} 

\thanks[footnoteinfo]{This work is financially supported by National Key R\&D Program of China (Grant 2022YFB4702400).}

\author[Past]{Liyong Lin}

\address[Past]{Contemporary Amperex Technology Limited (e-mail: llin5@e.ntu.edu.sg).}


\begin{abstract}     
Timed discrete-event systems (TDES), which is a modelling formalism proposed by Brandin and Wonham,  can be used for modelling scheduling and production planning problems. This paper aims to show that TDES are essentially synchronous product structures. The proof is constructive in the sense that a  generalized synchronous product rule is provided to generate a TDES from  the activity automaton and the timer automata (that is, the syntactic description of the TDES) after some model transformation. We then also explain how the generalized synchronous product operation can be reduced into the standard synchronous product operation and how to reduce the number of (refined) events introduced in the model transformation. Thus, any software that can compute synchronous products can be used to compute a TDES from its activity automaton and its timer automata, after the model transformation.  
\end{abstract}
\begin{keyword}                           
timed discrete-event systems, timed state-tree structures, synchronous product, supervisory control 
\end{keyword} 
\end{frontmatter}



\section{Introduction}
The theory of control for real-time discrete-event systems is a very rich subject that has been investigated in several different formalisms, e.g., the clock automata~\cite{BH88}, the timed transition models~\cite{90b}, the timed discrete-event systems~\cite{BW94}, with discrete time semantics, and the timed automata~\cite{TH91}, with dense time semantics. 
One of the most widely used models of real-time discrete-event systems in supervisory control theory is the timed discrete-event systems (TDES). TDES is   attractive since many existing techniques for the control of untimed discrete-event systems (DES) can often be naturally extended to this timed counterpart. Indeed, a few number of structured supervisor synthesis approaches have  been extended to the   TDES formalism (e.g. ~\cite{SQC17},~\cite{A07}). Some recent extensions of~\cite{BW94} include the supervisor localization~\cite{ZGWW13},~\cite{ZC19}, relative observability and relative coobservability~\cite{CZW16}, networked supervisor synthesis~\cite{ZCGW16},~\cite{RRF18},~\cite{PU20},  state-based control~\cite{RW18} and so on. Quite a few number of practical applications have been developed based on the framework of TDES (see, for example,~\cite{WS17},~\cite{ZLWZ17},~\cite{TC17} for some recent works) and interested reader may also see~\cite{S20} for an envisioned application of control of TDES. 
 
 The modeling framework of \cite{BW94}  starts with a (finite state) interval model, where each event is associated with a fixed discrete time interval. Then, the interval model has to be converted into a (finite state) tick model, which explicitly enumerates the {\em tick} events to model  time intervals in the original model. However, an inherent difficulty of TDES, compared with the untimed counterpart, is precisely due to the explicit enumeration of ticks that only makes the state explosion problem much worse, especially when the intervals have large upper bounds\footnote{An interval model could have an unbounded ``compression rate" compared with the (flat) tick model~\cite{LSB19}. For example, to simulate a transition in the interval model labeled by $(\sigma, [1, k])$, $k$ ticks-labeled transitions need to be created in the tick model.}.  A symbolic approach for the supervisory control of TDES has been successfully developed in~\cite{MFAL15}, by using the timed extended finite automata (TEFA) and binary decision diagrams (BDD). An approach that attempts to avoid an explicit enumeration of ticks can be found in~\cite{LSB19},~\cite{BSL20}, which operates on interval automata directly (by using interval arithmetic) for both synchronous product construction and supervisor synthesis. A timed state-tree structures (TSTS) based supervisor synthesis framework has  been developed in~\cite{A07} to cope with the state explosion issue encountered in TDES, encouraged by the success of the (untimed) state-tree structures based supervisor synthesis approach in~\cite{MW04}. However, the TSTS framework developed in~\cite{A07} seems far less successful, and only deals with  systems of state sizes of the order $\leq 10^{12}$, even if (the same) BDD based encoding of state trees is used. We conjecture one of the main reasons behind this is because the  modelling power of the state-tree structures (STS) formalism is not well utilized. Indeed, \cite{A07} uses each (monolithic) TDES as a holon, instead of using the (constituent) activity automaton and the timer automata as the holons. Thus, state explosion has implicitly occurred in the computation or model building of the TDES holons\footnote{We shall remark that this is not the only reason, as an explicit enumeration of ticks still cannot be avoided in the timer automaton holons. For example, if the intervals for events have large upper bounds, then each timer automaton holon can have a large state size. }. 
 
 In this paper, we shall show that this source of inefficiency of TSTS could be avoidable. Indeed, the aim of this paper is to show that each TDES of~\cite{BW94} is a synchronous product structure, which can be built from the activity automaton and the timer automata after some model transformation; thus, it is possible for one to use the activity automaton and the timer automata, instead of their product, that is, TDES, as the holons, which will hopefully help improve the efficiency of TSTS based supervisor synthesis procedure by increasing the horizontal modularity. This result may be of independent interest as well. We remark that, before this work, it is an open problem, to the best of our knowledge, whether a TDES can be built from its activity automaton and timer automata~\cite{BW94}. Indeed, it is stated in the second paragraph of Section V in~\cite{BW94} that ``Unfortunately there is no simple way to obtain $G$ by straightforward combination of $G_{act}$ with the $SPEC\sigma$". 

The paper is organized as follows. In Section 2, we review the basics of TDES. In Section 3, we show the main result that TDES are synchronous product structures. 
Finally, we provide conclusion and future works in Section 4.
\section{Basics of Timed Discrete-Event Systems}
In this section, we shall present some basics of the TDES formalism~\cite{BW94},~\cite{WMW17} to make this paper more self-contained. To that end, we need to first recall some general notation and terminology.

Let $\mathbb{N}$ denote the set $\{0, 1, 2, 3, \ldots\}$ of natural numbers. We write $[k_1, k_2]$, where $k_1, k_2 \in \mathbb{N}$ and $k_1 \leq k_2$, to denote the set $\{k \in \mathbb{N} \mid k_1 \leq k \leq k_2\}$, and write $[k_1, \infty)$ to denote the set $\{k \in \mathbb{N} \mid k \geq k_1\}$. Let $\mathcal{I}$ denote the collection of intervals of the above two types, and let $\mathcal{I}_f \subset I$ denote the collection of intervals of the first type. For any interval $I$, we define $I.l=k_1, I.r:=k_2$ if $I=[k_1, k_2]$; we define $I.l:=k_1, I.r:=\infty$ if $I=[k_1, \infty)$. For brevity, we use  $``t"$, instead of ``tick", to denote the tick event.
\subsection{Syntax}
The syntax of a TDES is given by a tuple $(G_{act}, T)$, where 
\begin{center}
$G_{act}=(A, \Sigma_{act}, \delta_{act}, a_0, A_m)$
\end{center}
is a finite state automaton (over $\Sigma_{act}$) and $T: \Sigma_{act} \mapsto \mathcal{I}$ maps each $\sigma \in \Sigma_{act}$ to an interval $T(\sigma) \in \mathcal{I}$. $G_{act}$ is  the {\em activity automaton}, where $A$ is the finite set of activities; $\Sigma_{act}$ is the finite set of (activity) events; $\delta_{act}: A \times \Sigma_{act} \mapsto A$ is the (partial) activity transition function\footnote{We also write $\delta_{act} \subseteq A \times \Sigma_{act} \times A$. As usual, we can extend $\delta_{act}$ to the partial function $\delta_{act}: A \times \Sigma_{act}^* \mapsto A$.}; $a_0 \in A$ is the initial activity and $A_m \subseteq A$ the subset of marker activities. $T$ is the {\em timer map} which naturally induces a partition of the event set $\Sigma_{act}=\Sigma_{spe} \dot{\cup} \Sigma_{rem}$, where\footnote{The subscript ``spe" denotes
``prospective", while  ``rem" denotes
``remote".} $\sigma \in \Sigma_{spe}$ if and only if $T(\sigma) \in \mathcal{I}_f$.  


\subsection{Semantics}
\label{sec: seman}
Recall that the semantics of the tuple $(G_{act}, T)$ is a finite state automaton $G=(Q, \Sigma, \delta, q_0, Q_m)$ over  $\Sigma:=\Sigma_{act} \dot{\cup} \{t\}$, often referred to as a timed discrete-event system (TDES). In the rest of this subsection, we shall explain how the four components $Q, \delta, q_0$ and $Q_m$ are generated from the tuple $(G_{act}, T)$. For an informal description of the semantics, the reader is referred to~\cite{BW94}.


 A timer interval $T_{\sigma}$ is defined for each event $\sigma$ as follows: 
\[   
T_{\sigma}:= 
     \begin{cases}
       [0, T(\sigma).r], & \text{if }\sigma \in \Sigma_{spe}\\
       [0, T(\sigma).l], & \text{if }\sigma \in \Sigma_{rem} \ 
     \end{cases}
\]
The state set is defined to be $Q:=A \times \prod_{\sigma \in \Sigma_{act}}T_{\sigma}$, where without loss of generality we arbitrarily fix an ordering in the enumeration of $\Sigma_{act}$ in the Cartesian product. A state is a tuple of the form $q=(a, (t_{\sigma})_{\sigma \in \Sigma_{act}})$, where $a \in A$ and, for each $\sigma \in \Sigma_{act}$, $t_{\sigma} \in T_{\sigma}$. The $t_{\sigma}$ component of $q$ is the timer value of $\sigma$ in $q$. The default timer value $t_{\sigma 0}$ for each $\sigma \in \Sigma_{act}$ is defined as follows:
\[   
t_{\sigma 0}:= 
     \begin{cases}
       T(\sigma).r, & \text{if }\sigma \in \Sigma_{spe}\\
       T(\sigma).l, & \text{if }\sigma \in \Sigma_{rem} \ 
     \end{cases}
\]
The initial state is  $q_0=(a_0, (t_{\sigma 0})_{\sigma \in \Sigma_{act}})$. And, the set of marker states is defined to be\footnote{In~\cite{BW94}, it is only required there that $Q_m \subseteq A_m \times \prod_{\sigma \in \Sigma_{act}}T_{\sigma}$. The software TTCT  provides two options: $Q_m=A_m \times \prod_{\sigma \in \Sigma_{act}}T_{\sigma}$ or $Q_m= A_m \times \prod_{\sigma \in \Sigma_{act}}\{t_{\sigma 0}\}$. The second option is adopted in this work, without loss of generality, as the marker states set can be easily adapted accordingly.}  $Q_m= A_m \times \prod_{\sigma \in \Sigma_{act}}\{t_{\sigma 0}\}$. The definition of the partial transition function $\delta$ is more tedious. Let $q=(a, (t_{\sigma})_{\sigma \in \Sigma_{act}})$ and $q'=(a', (t'_{\sigma})_{\sigma \in \Sigma_{act}})$ be any two states. We have the following three cases: 

\begin{enumerate}
\item [A)] for any $\sigma \in \Sigma_{spe}$, $\delta(q, \sigma)=q'$ if and only if 
\begin{enumerate}
\item  $\delta_{act}(a, \sigma)!$ and $0 \leq t_{\sigma} \leq T(\sigma).r-T(\sigma).l$
\item $a'=\delta_{act}(a, \sigma)$ and for each $\tau \in \Sigma_{act}$,
\begin{enumerate}
\item if $\tau \neq \sigma$, then \[   
t'_{\tau}:= 
     \begin{cases}
       t_{\tau 0}, & \text{if } \neg \delta_{act}(a', \tau)!\\
       t_{\tau}, & \text{if } \delta_{act}(a', \tau)! \ 
     \end{cases}
\]
\item if $\tau=\sigma$, then $t'_{\tau}=t_{\sigma 0}$
\end{enumerate}
\end{enumerate}
\item [B)] for any $\sigma \in \Sigma_{rem}$, $\delta(q, \sigma)=q'$ if and only if
\begin{enumerate}
\item $\delta_{act}(a, \sigma)!$ and $t_{\sigma}=0$
\item $a'=\delta_{act}(a, \sigma)$ and for each $\tau \in \Sigma_{act}$,
\begin{enumerate}
\item if $\tau \neq \sigma$, then \[   
t'_{\tau}:= 
     \begin{cases}
       t_{\tau 0}, & \text{if } \neg \delta_{act}(a', \tau)!\\
       t_{\tau}, & \text{if } \delta_{act}(a', \tau)! \ 
     \end{cases}
\]
\item if $\tau=\sigma$, then $t'_{\tau}=t_{\sigma 0}$
\end{enumerate}
\end{enumerate} 
\item [C)] $\delta(q, t)=q'$ if and only if
\begin{enumerate}
\item $\forall \tau \in \Sigma_{spe}, (\delta_{act}(a, \tau)! \implies t_{\tau}>0)$ 
\item $a'=a$ and for each $\tau \in \Sigma_{act}$,
\begin{enumerate}
\item if $\tau \in \Sigma_{spe}$, then \[   
t'_{\tau}:= 
     \begin{cases}
       t_{\tau 0}, & \text{if } \neg \delta_{act}(a, \tau)!\\
       t_{\tau}-1, & \text{if } \delta_{act}(a, \tau)! \wedge t_{\tau}>0\ 
     \end{cases}
\]
\item if $\tau \in \Sigma_{rem}$, then \[   
t'_{\tau}:= 
     \begin{cases}
       t_{\tau 0}, & \text{if } \neg \delta_{act}(a, \tau)!\\
       t_{\tau}-1, & \text{if } \delta_{act}(a, \tau)! \wedge t_{\tau}>0\\ 
       0, & \text{if } \delta_{act}(a, \tau)! \wedge t_{\tau}=0 \
     \end{cases}
\]
\end{enumerate}
\end{enumerate}
\end{enumerate}

We have now completed the description of the semantics of the tuple $(G_{act}, T)$ as a timed discrete-event system $G$. We finally remark that it is required that all TDES must be {\em activity-loop-free}, namely, $\forall q \in Q, s \in \Sigma_{act}^+, \delta(q, s)\neq q$.  

\section{TDES are Synchronous Product Structures}
We note that the description of the semantics of $(G_{act}, T)$ in Section~\ref{sec: seman} is a bit involved and may even be difficult to visualize (for beginners).  While all the other components of $G$ suggest that $G$ could be a synchronous product structure (for example, the state space $Q$ is a Cartesian product), the definition of $\delta$ is not presented in an explicitly structured manner nor is it clear that $\delta$ has a product structure; and, it is difficult to obtain much insight from the definition of $\delta$. A structured construction of $G$ by combining\footnote{$T$ is only a timer map here; thus, it is expected that we need to transform $T$ into some automata. The details will be presented soon.} $G_{act}$ and $T$ is an open problem that still remains not addressed~\cite{BW94}. In the following, we show that a generalized synchronous product operation for the construction of a TDES from its activity automaton and timer map description, after some model transformation, is feasible. Later, we also show how the generalized product operation can be mapped into the standard synchronous product operation. Furthermore, we will show that the synchronous product based construction exactly matches the construction provided in Section~\ref{sec: seman}. 

The overall construction consists of four steps; the details will be explained in the rest of this section. 
\subsection{Step 1: Timer Map to Timer Automata}
In order for the synchronous product to be computed, the first step is to transform the timer map to a collection of timer automata. For each pair $(\sigma, T(\sigma))$, we shall define a timer automaton $G_{\sigma}=(Q_{\sigma}, \Sigma_{\sigma}, \delta_{\sigma}, q_{0, \sigma}, Q_{m, \sigma})$. Before we provide the formal definition of $G_{\sigma}$, we shall use the next example as an illustration. 



The formal construction is as follows.

1) If $\sigma \in \Sigma_{spe}$, then we let $Q_{\sigma}=[0, T(\sigma).r]$, $\Sigma_{\sigma}=\{t, \sigma\}$, $q_{0, \sigma}=T(\sigma).r$ and $Q_{m, \sigma}=\{T(\sigma).r\}$. $\delta_{\sigma}$ is specified by its graph\footnote{The graph of a  partial function $f$ with domain $Dom$ is the relation $\{(x, f(x)) \mid x \in Dom\}$.}, which is the union of $\{(i+1, t, i) \mid i \in [0, T(\sigma).r-1]\}$ and $\{(i, \sigma, T(\sigma).r) \mid i \in [0, T(\sigma).r-T(\sigma).l]\}$. 

2) If $\sigma \in \Sigma_{rem}$, then we let $Q_{\sigma}=[0, T(\sigma).l]$, $\Sigma_{\sigma}=\{t, \sigma\}$, $q_{0, \sigma}=T(\sigma).l$ and $Q_{m, \sigma}=\{T(\sigma).l\}$. $\delta_{\sigma}$ is specified by its graph, which is the union of $\{(i+1, t, i) \mid i \in [0, T(\sigma).l-1]\}$ and $\{(0, \sigma, T(\sigma).l), (0, t, 0)\}$. 


Each timer automaton $G_{\sigma}$ constructed above corresponds to the specification $SPEC\sigma$ over  $\{t, \sigma\}$ in Section V of~\cite{BW94}. Despite of the fact that the timer automata have already been constructed in~\cite{BW94}, it is not known whether and how they could be used in the computation of $G$ in~\cite{BW94}. In the following subsections, we  provide the remaining three steps to complete the construction of $G$ by using a  synchronous product operation, after some model transformation.
\subsection{Step 2: Automata Transformation}
The automata $(G_{act}, (G_{\sigma})_{\sigma \in \Sigma_{act}})$ cannot be used for synchronous product. The main idea to resolve the difficulties is to 1) add labels to tick transitions and event transitions in the activity automaton to reflect, respectively, the status of enablement of events and the effect of event transitions on the enablement of events, and 2) split and add labels to tick transitions and event transitions in the timer automata to reflect the effects of different transitions\footnote{Here, event transitions refer to those transitions labeled by events in $\Sigma_{act}$.}. 

Let $\Sigma_{act}=\{\sigma_1, \sigma_2, \ldots, \sigma_n\}$. The transformation can then be summarized as follows:\\
{\bf [Computing Transformed Activity Automaton]}: \\
Given $G_{act}$ (before adding the self-loops), we perform the following operations.
\begin{enumerate}
\item for each state $a \in A$ in $G_{act}$, add a self-loop labeled by $(t, ph_1, ph_2, \ldots, ph_n)$, where $ph_i$ is a place holder that is to be replaced by $\sigma_i!$ if $\sigma_i$ is defined at state $a$, and replaced by $\neg \sigma_i!$ if $\sigma_i$ is not defined at state $a$.
\item for each transition $(a, \sigma, a')$, where $\sigma \in \Sigma_{act}$, replace it with the transition $(a, (\sigma, ph_1', ph_2', \ldots, ph_n'), a')$, i.e., add the label $(ph_1', ph_2', \ldots, ph_n')$ to $\sigma$ in the transition $(a, \sigma, a')$. Here, $ph_i'$ is a place holder that is to be replaced by $E \sigma_i$ if $\sigma_i$ is defined at state $a'$, and replaced by $D \sigma_i$ if $\sigma_i$ is not defined at state $a'$.
\end{enumerate}

Intuitively, the meaning of $(t, \alpha !, \neg \beta !)$ at state $0 \in A$ is that $\alpha$ is enabled and $\beta$ is disabled at state $0 \in A$. The meaning of the transition $(0, (\alpha, D\alpha, E\beta), 1)$ is that after firing event $\alpha$ at state 0, $\alpha$ is disabled and $\beta$ is enabled (in the next state). The other cases can be explained in a similar way. 


We shall use the next example to illustrate the transformation procedure for the timer automata.

The transformation can be summarized as follows:\\
{\bf [Computing Transformed Timer Automaton]}: \\
Given $G_{\sigma}$ (before adding the self-loops), we perform the following operations. Recall that \[   
t_{\sigma 0}:= 
     \begin{cases}
       T(\sigma).r & \text{if }\sigma \in \Sigma_{spe}\\
       T(\sigma).l & \text{if }\sigma \in \Sigma_{rem} \ 
     \end{cases}
\]
is the initial state of $G_{\sigma}$.
\begin{enumerate}
\item for each state $i \in Q_{\sigma}$ and each event $\sigma' \in \Sigma_{act}$ other than $\sigma$, add the transition $(i, (\sigma', E\sigma), i)$, i.e., add a self-loop labeled by $(\sigma', E\sigma)$, and add the transition $(i, (\sigma', D\sigma), t_{\sigma 0})$.
\item for each transition $(i, \sigma, t_{\sigma 0})$, replace it with the two transitions $(i, (\sigma, E\sigma), t_{\sigma 0})$ and $(i, (\sigma, D\sigma), t_{\sigma 0})$.
\item for each transition labeled by $t$, replace the label with $(t, \sigma !)$.
\item for the state $t_{\sigma 0} \in Q_{\sigma}$, add the transition $(t_{\sigma 0}, (t, \neg \sigma !), $\\$t_{\sigma 0})$,  i.e., add a self-loop labeled by $(t, \neg \sigma !)$.
\end{enumerate}

Intuitively, labels are added to reflect the effects of different transitions to the timer value for each timer automaton. In the above procedure, we split tick transitions and event transitions for exactly that purpose. 

\subsection{Step 3: Generalized Synchronous Product}
Up to now, we have completed the transformation of the activity automaton and the timer automata. It is straightforward to define a generalized synchronous product operation that combines the transformed activity automaton and the transformed timer automata. The synchronization constructs are shown in the following.
\begin{enumerate}
\item for event $(t, ph_1, ph_2, \ldots, ph_n)$ in the transformed activity automaton, it can be synchronized with event $(t, ph_i)$ in the transformed timer automaton for $\sigma_i$, for $i \in [1, n]$, where $ph_i$ is a place holder for $\sigma_i !$ or $\neg \sigma_i !$; after the synchronization, the label is event $t$  
\item for event $(\sigma, ph_1', ph_2', \ldots, ph_n')$ in the transformed activity automaton,  where $\sigma \in \Sigma_{act}$, it can be synchronized with event $(\sigma, ph_i')$ in the transformed timer automaton for $\sigma_i$, for $i \in [1, n]$, where $ph_i'$ is a place holder for $E\sigma_i$ or $D\sigma_i$; after the synchronization, the label is event $\sigma$ 
\end{enumerate}

Let $G_{act}^T$ denote the transformed activity automaton and $G_{\sigma}^T$ the transformed timer automaton for $\sigma \in \Sigma_{act}$. If we use $\bowtie$ to denote the generalized synchronous product operation, then $G_{act}^T \bowtie (\bowtie_{\sigma \in \Sigma_{act}} G_{\sigma}^T)$ can be constructed using the above synchronization constructs.

\subsection{Step 4: Generalized Synchronous Product to Standard Synchronous Product}

To map the generalized synchronous product operation $\bowtie$ into the standard synchronous product operation $\lVert$, the last step involves straightforward relabelling for each timer automaton. Thus, for each timer automaton, we perform the following.
\begin{enumerate}
    \item replace each transition that is labeled with $(\sigma, ph_i)$ with the set of transitions labeled with \begin{center}$\{(\sigma, \overline{ph_1},\ldots, \overline{ph_{i-1}},ph_i, \overline{ph_n}) \mid \overline{ph_j}=E\sigma_j \vee \overline{ph_j}=D\sigma_j, j \in [1, n]-\{i\}\}$
    \end{center}
    \item replace each transition that is labeled with $(t, ph_i)$ 
    with the set of transitions labeled with \begin{center}$\{(t, \overline{ph_1},\ldots, \overline{ph_{i-1}},ph_i, \overline{ph_n}) \mid \overline{ph_j}=\sigma_j! \vee \overline{ph_j}=\neg \sigma_j!, j \in [1, n]-\{i\}\}$
    \end{center}
\end{enumerate}
Let $R(G_{\sigma}^T)$ denote the relabelled transformed timer automaton, for each $\sigma \in \Sigma$. Then, we can obtain the synchronous product $G_{act}^T \lVert (\lVert_{\sigma \in \Sigma_{act}}R(G_{\sigma}^T))$. Each event of $G_{act}^T \lVert (\lVert_{\sigma \in \Sigma_{act}}R(G_{\sigma}^T))$ is of the form $(\sigma, ph_1, ph_2,\ldots, ph_n)$ or $(t, ph_1', ph_2', \ldots, ph_n')$, which contains refined event information. It is  straightforward to recover the original event information by hiding the placeholder information, by replacing $(\sigma, ph_1, ph_2,\ldots, ph_n)$ with $\sigma$ and $(t, ph_1', ph_2', \ldots, ph_n')$ with $t$. We denote the resulting finite state automaton $h(G_{act}^T \lVert (\lVert_{\sigma \in \Sigma_{act}}R(G_{\sigma}^T)))$.

It appears that an exponential number of events need to be created for the model transformation, to be used in the synchronous product constructions. This can be easily avoided by relabeling the events in $G_{act}$ first. Then, one only need to transform each $G_{\sigma}$ with the events used in $G_{act}^{T}$. Thus, the number of events used is upper bounded by the number of transitions of $G_{act}$.

\section{Conclusions and Future Work}
In this work, we have shown that TDES are synchronous product structures, thus resolving a  problem from~\cite{BW94} that has been unaddressed. Moreover, the technique presented in this work opens up some new directions for future research. An interesting question is whether a modular presentation of TDES can already be useful for the modular control. This work can be viewed as reducing the semantic interpretation of TDES into the computation of the synchronous products. We will present a sequel of this work "Reduction for Time Discrete-Event Systems II: From Untimed Synthesis to Timed Synthesis and Back" in a companion paper, and present the application of control~\cite{BSL20},\cite{ZLTS22},\cite{BSL24} for scheduling and production planning. 

{\bf Acknowledgement}
The first author would like to thank Sadegh Rahnamoon and Prof. Wonham for commenting on an earlier version of this paper.


\begin{thebibliography}{}
\bibitem[Brave(1988)]{BH88}
Y.~Brave, M.~Heymann. {``Formulation and control of real time discrete event processes"}, CDC, pp. 1131–1132, 1988.


\bibitem[Ostroff(1990)]{90b}
J.~S.~Ostroff. {``A framework for real-time discrete-event control"}, IEEE Transactions on Automatic Control, 35(4): 386-397, 1990.

\bibitem[Brandin(1994)]{BW94}
B.~A.~Brandin, W.~M.~Wonham. {``Supervisory control of timed discrete-event systems"}, IEEE Transactions on Automatic Control, 39(2): 329-342, 1994.

\bibitem[Toi(1991)]{TH91}
H.~Wong-Toi, G.~Hoffmann. {``The control of dense real-time discrete event systems"}, Proceedings of the 30th IEEE Conference on Decision and Control, 1527-1528, 1991.

\bibitem[Schafaschek(2017)]{SQC17}
G.~Schafaschek, M.~H. de Queiroz, J.~E. R. Cury. {``Local modular supervisory control
of timed discrete-event systems"}, IEEE Transactions on Automatic Control, 62(2): 934-940, 2017.

\bibitem[Saadatpoor(2009)]{A07}
Ali Saadatpoor. \emph{Timed state tree structures: supervisory control and fault diagnosis}, PhD Thesis, University of Toronto, Toronto, Canada, 2009.


\bibitem[Zhang(2013)]{ZGWW13}
R. Zhang, K. Cai, Y. Gan, Z. Wang, W. M. Wonham.
{``Supervision localization of timed discrete-event systems"},
Automatica, 49(9): 2786-2794, 2013.

\bibitem[Ostroff(2019)]{ZC19}
R. Zhang, K. Cai. {``Supervisor localization of timed discrete-event systems under partial observation"}, IEEE Transactions on Automatic Control, DOI: 10.1109/TAC.2019.2912008, 2019.


\bibitem[Cai(2016)]{CZW16}
K. Cai, R. Zhang, W.M. Wonham. {``Relative observability
and coobservability of timed discrete-event systems"}, IEEE
Transactions on Automatic Control, 61(11): 3382-3395, 2016.

\bibitem[Zhang(2016)]{ZCGW16} R. Zhang, K. Cai, Y. Gan, W.M. Wonham. {``Delay-robustness
in distributed control of timed discrete-event systems based
on supervisor localization"}, International Journal of Control,
89(10): 2055-2072, 2016.


\bibitem[Rashidinejad(2018)]{RRF18}
A. Rashidinejad, M. Reniers, L. Feng. {``Supervisory control of timed discrete-event systems subject to communication delays and non-FIFO observations"},  Workshop on Discrete Event Systems, pp. 456-463, Sorrento, 2018.

\bibitem[Pruekprasert(2020)]{PU20}
S.  Pruekprasert,  T.  Ushio.  {``Supervisory  control  of communicating timed discrete event systems for state avoidance  problem"}, IEEE Control Systems  Letters,4(1):259-264, 2020.


\bibitem[Rahnamoon(2018)]{RW18} S. Rahnamoon, W. M. Wonham. {``State-based control of timed discrete-event systems"}, In Proc. 2018 IEEE Conference  on  Decision  and  Control, pp. 4833-4838, Miami, 2018.

\bibitem[Ware(2017)]{WS17}
 S. Ware, R. Su. {``Time optimal synthesis based upon sequential  abstraction  and  its  application  in  cluster tools"}, IEEE Transactions on Automation Science and Engineering, 14(2):772-784, 2017.

\bibitem[Zhao(2017)]{ZLWZ17}
  B.  Zhao,  F.  Lin,  C.  Wang,  X.  Zhang,  M.  P.  Polis, L. Y. Wang. {``Supervisory control of networked timed discrete event systems and its applications to power distribution networks"}, IEEE Transactions on Control of Network Systems, 4(2):146-158, 2017.
  
 \bibitem[Monteiro(2017)]{TC17}
 T. Monteiro Tuxi, A. Carrilho da Cunha, 
 {``Timed supervisory control of an industrial glass bonding system"}, 50(1): 12339-12344, 2017.

\bibitem[Seow(2020)]{S20}
K. T. Seow, Supervisory control of blockchain networks, IEEE Transactions on Systems, Man, and Cybernetics: Systems, 50(1): 159-171, 2020.

\bibitem[Lin(2019)]{LSB19}
L. Lin, R. Su, B. A. Brandin, S. Ware, Y. Zhu, Y. Sun. {``Synchronous composition of finite interval automata"}, IEEE International Conference on Control and Automation, pp. 578-583, 2019.

\bibitem[Miremadi(2015)]{MFAL15}
S. Miremadi, Z. Fei , K. Åkesson , B. Lennartson, 
{``Symbolic Supervisory Control of Timed Discrete Event Systems"}, IEEE Transactions on Control Systems Technology, 584-597, 2015.

\bibitem[Ma(2004)]{MW04}
C.~Ma, W.~M.~Wonham. \emph{Nonblocking supervisory control of state tree structures}, Springer, 2004.

\bibitem[Wonham(2021)]{WMW17}
W.~M.~Wonham, K.~Cai. \emph{Supervisory control of discrete-event systems}, Springer, 2021.

\bibitem[Brandin(2020)]{BSL20}
B. A. Brandin, R. Su,  L. Lin. {``Supervisory control of time-interval discrete-event systems"},  Workshop on Discrete Event Systems, IFAC-PapersOnline 53(4), 217-222, 2020.

\bibitem[Zhu(2020)]{ZLTS22}
Y. Zhu, L. Lin, R. Tai, R. Su.  {``
Distributed Control of Timed Networked System against Communication Delays"}, ICCA, 1008-1013, 2022.

\bibitem[Brandin(2024)]{BSL24}
B. A. Brandin, R. Su, L. Lin:
{``Supervisory Control of Time-Interval Discrete Event Systems"}, IEEE Trans. Autom. Control, 69(5): 3080-3095, 2024.

\end{thebibliography}
\end{document}